\documentclass[10pt]{iopart}
\usepackage{iopams}
\usepackage{graphicx,epsfig}
\usepackage{dcolumn}
\usepackage{bm}

\def\fun#1#2{\lower3.6pt\vbox{\baselineskip0pt\lineskip.9pt
\ialign{$\mathsurround=0pt#1\hfil##\hfil$\crcr#2\crcr\sim\crcr}}}
\def\lap{\mathrel{\mathpalette\fun <}}

\newcommand{\be}{\begin{equation}}
\newcommand{\ee}{\end{equation}}
\newcommand{\bq}{\begin{eqnarray}}
\newcommand{\eq}{\end{eqnarray}}
\newcommand{\bc}{\begin{center}}
\newcommand{\ec}{\end{center}}

\begin{document}

\title[Consistency among distance measurements]{Consistency among distance measurements:  transparency,  BAO scale and accelerated  expansion}

\author{Anastasios Avgoustidis$^{1,2}$, Licia Verde$^{3,1}$, Raul Jimenez$^{3,1}$}
{\it $^1$ Institute of Space Sciences (IEEC-CSIC), Faculty of Sciences, Campus UAB, Bellaterra, Spain--tasos@ecm.ub.es\\
$^2$ Department of Structure and Constituents of Matter,  Physics, University of Barcelona, Spain\\
$^3$ Institucio Catalana de Recerca i Estudis Avancats (ICREA), 23 Passeig Lluis Companys, Barcelona, Spain--verde,raul@ieec.uab.es\\}

\date{\today}

\begin{abstract}
We explore consistency among different distance measures, including
Supernovae Type Ia data, measurements of the Hubble parameter, and
determination of the Baryon acoustic oscillation scale. We present new
constraints on the cosmic transparency combining $H(z)$ data together with the
latest Supernovae Type Ia data compilation. This combination, in the context of
a flat $\Lambda$CDM model, improves current constraints by nearly an order of
magnitude  although the constraints presented here are parametric rather than
non-parametric. We re-examine the recently reported tension between the Baryon
acoustic oscillation scale and Supernovae data in light of possible deviations
from transparency, concluding that the source of the discrepancy may most
likely be found among systematic effects of the modelling of the low redshift
data or a simple $\sim 2$-$\sigma$ statistical fluke, rather than in exotic
physics. Finally, we attempt to draw model-independent conclusions about the
recent accelerated expansion, determining the acceleration redshift to be
$z_{acc}=0.35^{+0.20}_{-0.13}$ (1-$\sigma$).
\end{abstract}



\section{Introduction}
In the -extremely successful- standard LCDM model there are well-defined relationships between different distance measures such as luminosity distance, angular diameter distance, BAO scale, etc.  While combining these measurements helps breaking parameter degeneracies and constraining cosmological parameters,  comparing them helps constraining possible deviations from the assumptions underlying the standard cosmological model or exotic physics. This is what we  set out to do here.

We will first consider  cosmic ``transparency".
The Etherington relation \cite{Etherington1} provides a distance duality which holds  for general metric theories of gravity in any background:
conservation of phase-space photon density and Lorentz-invariance implies that the luminosity distance  is $(1+z)^2$ times the angular diameter distance, where $z$ denotes the redshift.  This duality can be a powerful test  of exotic physics as well as simpler astrophysical effects. For example, it depends
crucially on photon conservation which can be violated by any source of attenuation.
 There are several known sources of attenuation, which are expected to be clustered e.g., interstellar dust, gas and plasma around galaxies. More speculatively,  if dark matter  is axion or axion-like it can have interactions with photons and produce effective absorption. The effect of clustered sources of attenuation, which are then expected to be clustered with the large-scale structures,  can be constrained with ``angular difference" measurements  and/or with cross correlation with large-scale structures  \cite{Menard,Bovy} and are at the 0.1\% level. In principle, there could also be an unclustered component of attenuation, which, however, is much more difficult to constrain.
 For example, there have been several attempts to explain the observed dimming of Type Ia Supernovae  without resorting to cosmic acceleration, but invoking unclustered sources of attenuation such as photon-axion mixing  \cite{Csaki1,Csaki2,MirRafSerp, Burrage} or gray dust \cite{Aguirre}. Both of these effects would have to have little frequency dependence above microwave frequencies to evade detection through reddening while still contributing to overall opacity\cite{Mortsell1,Mortsell2}.

 Radial and angular distance measures  can be combined to constrain this opacity \cite{More, BassettKunz1,BassettKunz2}.  In particular, \cite{More} combine Baryon Acoustic Oscillations (BAO) and Supernovae type Ia data to  constrain the difference in opacity between two different redshifts using present data and forecast  constraints from  forthcoming BAO measurements. Here we  build upon  \cite{More}: we combine Supernovae data with independent determinations of the cosmic expansion history $H(z)$ \cite{SVJ05}  to improve constraints on possible deviations from transparency.

Within the $\Lambda$CDM model, this data set combination, even allowing for opacity, yields predictions for the  observed  BAO scale.
We next compare these predictions  with the measured values at $z=0.35$ and $z=0.2$ \cite{EisensteinBAO, PercivalBAO}.  \cite{PercivalBAO} has hinted at a 2.4-$\sigma$ discrepancy between distance measures derived from Supernovae and BAO within the $\Lambda$CDM model. We revisit here this issue comparing distance measures derived from CMB \cite{wmap5}, from BAO  \cite{EisensteinBAO, PercivalBAO}, and from our own transparency analysis.

Finally, after exploring  the complementarity and consistency of different distance measures, we attempt to draw some model-independent conclusions about the recent  accelerated expansion by combining independent data-sets that are highly consistent with each other.
In particular, combining expansion history \cite{JVST,SVJ05}, SNIa \cite{MortsellClarkson} and HST
key \cite{HSTKey} data, we find evidence for recent acceleration at $z<0.5$ and indications of  past
deceleration at $z>1$.  We determine the transition redshift from deceleration to acceleration to be $z_{acc}=0.35^{+0.20}_{-0.13}$ at the $1$-$\sigma$ level.

\section{Transparency}
Luminosity distance and angular diameter distance are related by the
``Etherington relation'' \cite{Etherington1}:
\begin{equation}
d_L(z)=(1+z)^2 d_A(z) \,.
\label{Etherington}
\end{equation}
This relation holds  for general metric theories of gravity in
any background; it depends only on conservation of phase-space density of photons (transparency) and Lorentz invariance. Models that violate Lorentz invariance have non-trivial dispersion relations in the vacuum. This effect however generally becomes larger with increasing energy and is expected to be immeasurably small at low-energy (visible bands).
Transparency, however, could be violated by any source of photon attenuation. While  angular difference measurements \cite{Menard,Bovy} have imposed strong constraints on clustered opacity (i.e. strong upper limits on differences in opacity along different lines of sight), there could still be an unclustered source of attenuation, which is much harder to constrain.
Examples of effects that appear as uniform  attenuation include, gray dust \cite{Aguirre},  replenishing dust \cite{Riess04} (used to try to explain Supernovae results without resorting to cosmic acceleration, or to relax
the constraints on cosmic acceleration) or more exotic physics such as  axion-photon mixing \cite{Csaki1,Csaki2}.
\cite{BassettKunz1} propose to combine Supernovae measurements of $d_L$ with various estimates of $d_A$ to constrain exotic physics violating Etherington's distance duality.
\cite{More} propose to combine BAO and SN Ia data and  constrain the difference in opacity between two different redshifts ($z=0.2$ and $z=0.35$) from current data.

In this section we  develop this approach further and use $H(z)$ data \cite{JVST,SVJ05} in combination with Supernovae data \cite{Union} to constrain possible deviations from the ``Etherington relation''.
 If there was a source of ``photon  absorption" affecting the universe transparency, the distance modulus   derived from Supernovae would be systematically affected, in particular any effect that reduce the number of photons would dim the Supernovae brightness and increase $d_L$.  Let $\tau(z)$ denote the opacity  between an observer at $z=0$ and a source at $z$ due to e.g. extinction.   The flux received from the source would be attenuated by a factor $\exp(-\tau(z))$ and thus for the luminosity distance\footnote{If one wanted to interpret the optical depth  in terms of the comoving  number density of absorbers $n(z)$ and their cross section $\sigma(z)$ it will be
 \begin{equation}
 \tau(z)=\int_0^zn(z)\sigma(z)c \frac{(1+z)^2}{H(z)} dz=\int_0^zn(z)\sigma(z)\frac{c}{H_0} \frac{(1+z)^2}{E(z)} dz\,.
 \end{equation}}
 \begin{equation}
 d_{L, obs}^2=d_{L,true}^2 e^{\tau(z)} \, .
 \end{equation}
 Therefore, the inferred (``observed") distance modulus would be:
 \begin{equation}
 DM_{obs}(z)=DM_{true}(z)+2.5[\log e] \tau(z) \, .
 \end{equation}
Measurements of $DM_{obs}$ are taken from  the Supernovae Union compilation \cite{Union}.
We shall compare this with the -unabsorbed- luminosity distance $d_{L,true}$ inferred from the $H(z)$ measurements of \cite{SVJ05}. This measurement  is obtained from ages of old passively evolving galaxies: it relies on the detailed shape of the galaxy spectra but not on the galaxy luminosity. It will therefore not be affected by a non-zero  $\tau(z)$ since $\tau$ is assumed (and constrained by independent observations \cite{Mortsell1,Mortsell2}) not to be  strongly wavelength dependent in the optical band.

In particular, for a general FLRW cosmology:
\begin{equation}
 d_{L,true}(z)=(1+z)\frac{c}{H_0}\frac{1}{\sqrt{\Omega_k}}Sk(\sqrt{\Omega_k}\int_o^z\frac{dz'}{E(z)})
 \end{equation}
 where $Sk(x)$ stands for $sin(x)$, $x$, or $sinh(x)$ depending on $\Omega_k$ being positive, zero or negative respectively,
  and
  \begin{equation}
  E(z)=H(z)/H_0=[\Omega_m(1+z)^3+\Omega_V (1+z)^{3(1+w)}]^{1/2}\,.
  \end{equation}

  To be able to use the full redshift range of the available data, we consider the following simple parameterization of a deviation from the Etherington relation  $d_L=d_A(1+z)^{(2+\epsilon)}$, with $\epsilon$ parameterizing departures from transparency.
To understand the physical meaning of a constraint on $\epsilon$   we note that for small $\epsilon$ and $z \lap 1$ this is equivalent to assuming an optical depth parameterization  $\tau(z)=2\epsilon z$ or  $\tau=(1+z)^{\alpha}-1$ with the correspondence $\alpha=2 \epsilon$. While this identification is based on a Taylor expansion we have verified that the expansion is good to better than $20\%$ for the entire range  of $\epsilon$ and the redshift range considered and, as explained below,  to better than 1\% in the allowed rage.  
We consider a flat $\Lambda$CDM underlying  model: $H(z)$ and thus $d_{L,true}$  depends only on $\Omega_m$, $H_0$, while $d_{L,obs}$ and thus $DM_{obs}$ depends on $\Omega_m$, $H_0$ and $\epsilon$.  We fit the two data sets \cite{SVJ05} and \cite{Union}  both separately and jointly, imposing a Hubble constant prior from the HST key project \cite{HSTKey}.  In effect, the HST key project introduces one extra data point at low redshift.

\begin{figure}[h]
  \begin{center}
    \includegraphics[width=\columnwidth,keepaspectratio]{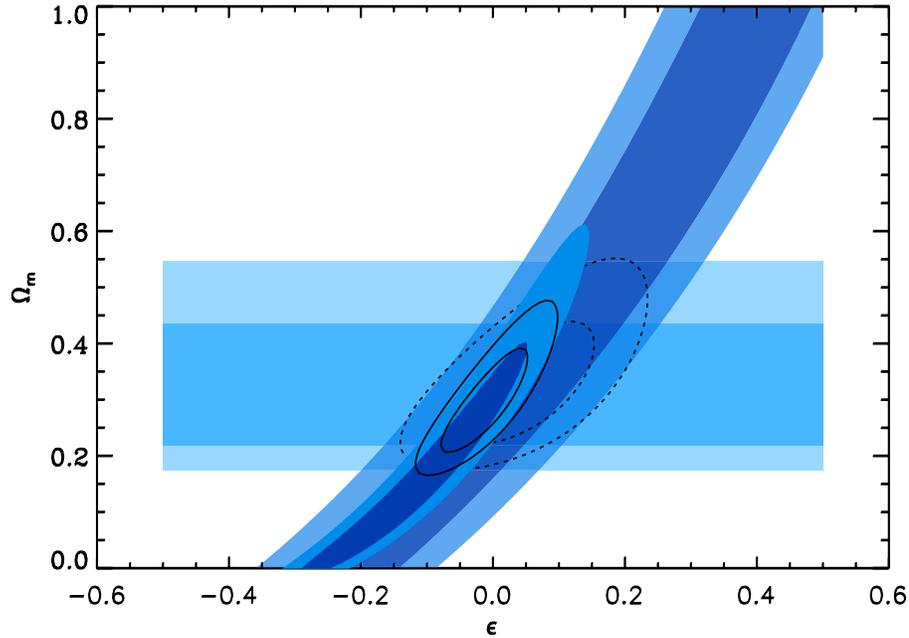}
  \end{center}
  \caption{\label{tau_omm_constr}  Confidence levels (68\% and 95\%) in the $\Omega_m$,$\epsilon$  plane from the joint analysis of Supernovae and $H(z)$ data, after marginalization over $H_0$. Dark blue regions are constraints from Supernovae data (contours in the background use only data at $z<0.5$ while contours in the foreground use all data), light blue regions from $H(z)$ data, and black transparent contours are the combined constraints. Here we have assumed an underlying flat  $\Lambda CDM$ model, where Supernovae-inferred luminosity distances are affected by a deviation from the Etherington relation parameterized by $\epsilon$. }
 \end{figure}

Figure \ref{tau_omm_constr} shows the constraints in the $\Omega_m$,$\epsilon$ plane after
marginalization\footnote{This should not be confused with maximization: we do not choose the values of $H_0$ that maximize the likelihood but actually integrate the likelihood over  the $H_0$.} over $H_0$. The dark blue regions are the $1$ and $2$-$\sigma$ joint (2-parameters) confidence levels for the Supernovae data. The dark blue contours in the background show the constraints considering only Supernovae at $z<0$ while the dark blue contours in the foreground correspond to all redshifts  . With the  extra degree of freedom of opacity, Supernovae data can only impose an upper limit to $\Omega_m$ even if the universe is assumed to be spatially flat. Note that negative values of $\epsilon$ would correspond to Supernovae being brighter than expected, which is unphysical if interpreted in terms of transparency, but it can still be interpreted in terms of departures from  the Etherington relation. The supernova-only constraints do not extend to arbitrarily large values of $\epsilon$. This is due to the presence of high redshift  data  breaking the degeneracy: the high redhsift supernova data show that the Supernovae dimming is evident at low redshift but not at high redshift \cite{Union,PerShaf}.
The lighter blue regions show the corresponding constraints from $H(z)$ data. As the $H(z)$ measurement is not affected by transparency, this data set yields no constraints on $\epsilon$. The dotted and  solid black lines show the combined constraints for the case of Supernovae data only at $z<0.5$ and for all redshifts respectively.

\begin{figure}[h]
  \begin{center}
    \includegraphics[width=0.45\columnwidth,keepaspectratio]{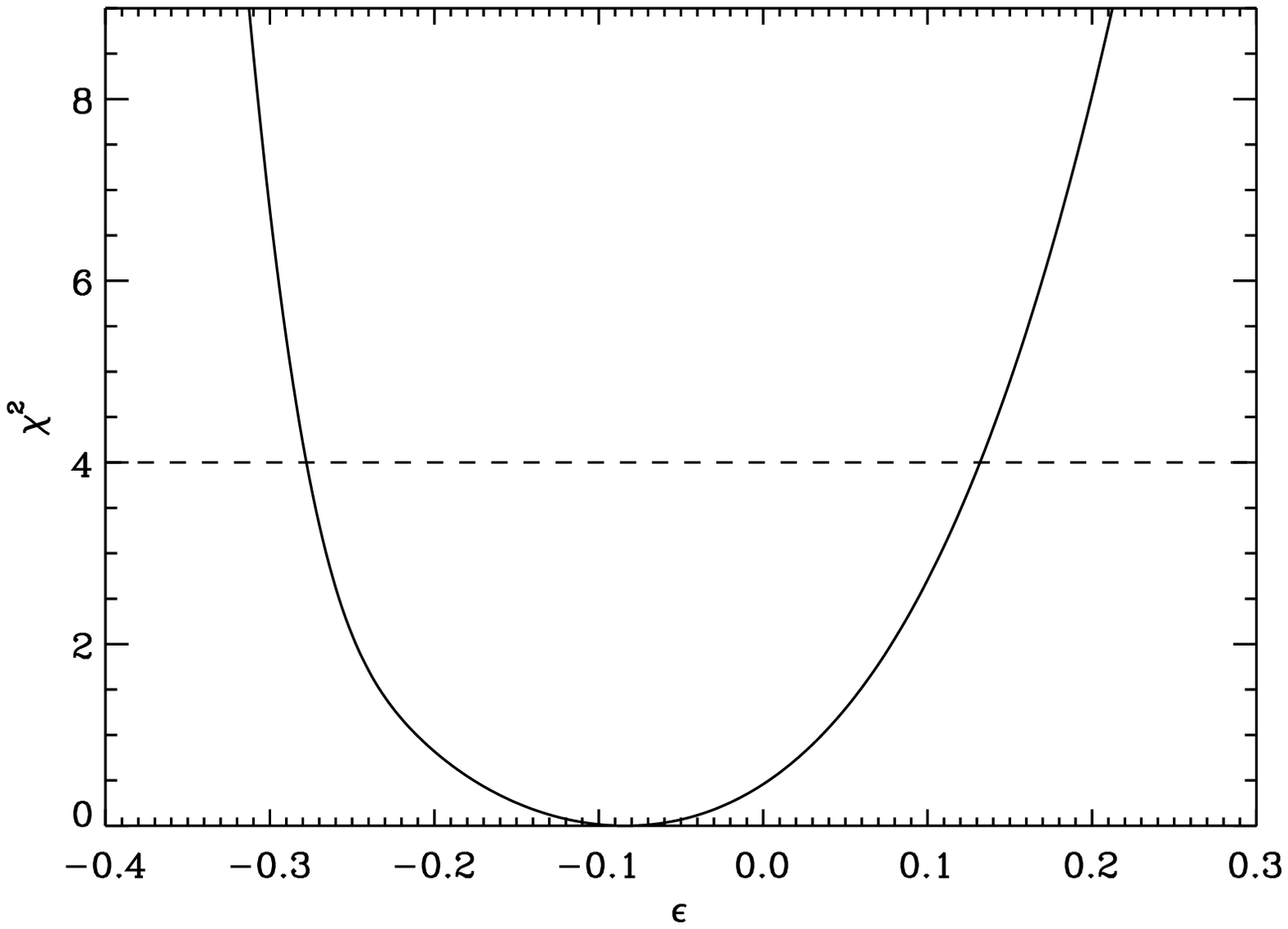}
    \includegraphics[width=0.45\columnwidth,keepaspectratio]{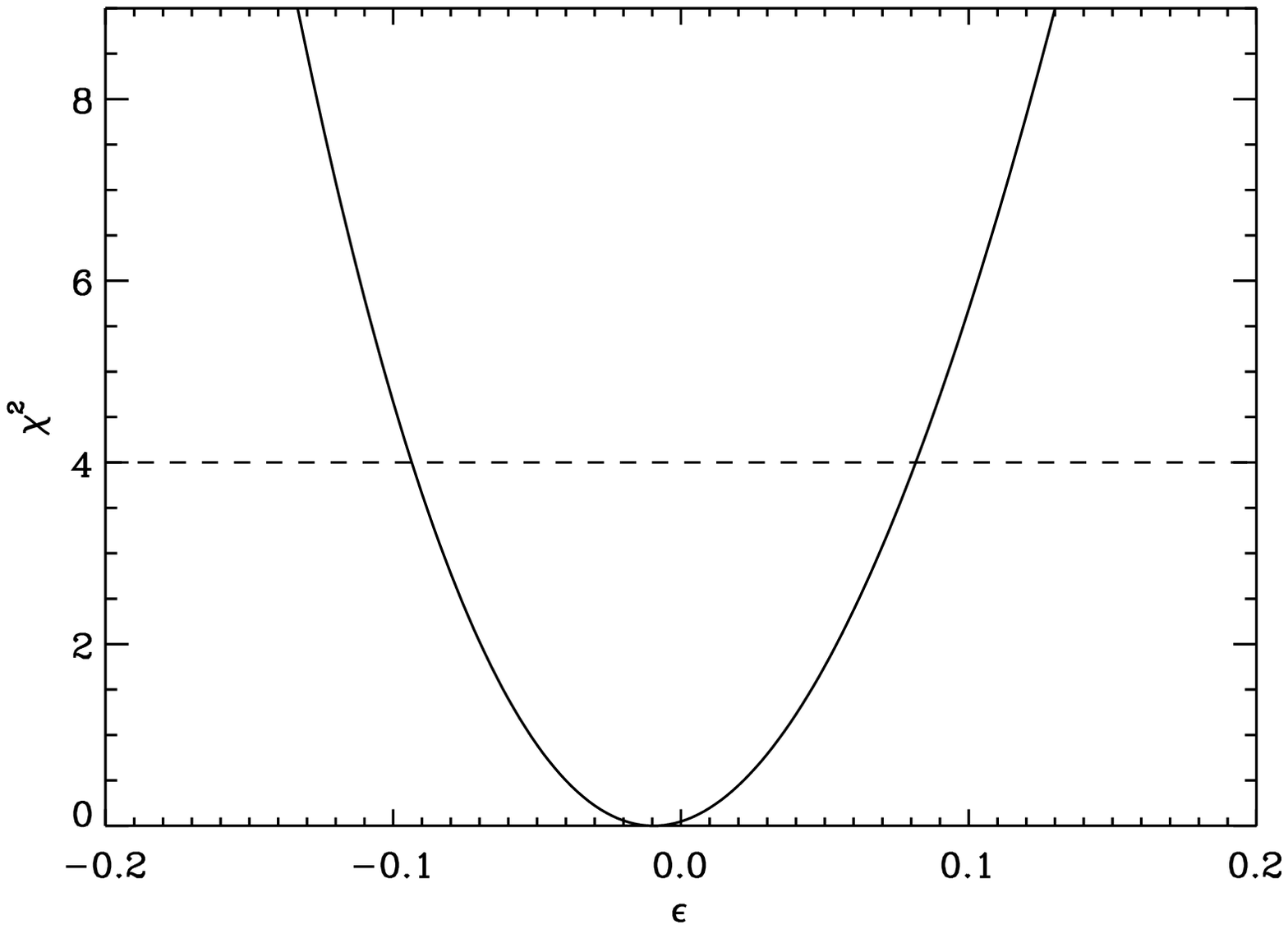}
  \end{center}
  \caption{\label{tau_constr} $\chi^2$ for $\epsilon$ (having marginalised over
  $\Omega_m$ and $H_0$) in the case of SN data only (left panel) and using both datasets
  (right panel). The dashed line corresponds to 95\% CL, $\Delta\chi^2=4$.}
 \end{figure}

Figure ~\ref{tau_constr}  shows the constraints on $\epsilon$ after marginalization over the other parameters. We obtain  $\epsilon=-0.08^{+0.21}_{-0.20}$  for Supernovae data alone (all redshifts) and  $\epsilon=-0.01^{+0.08}_{-0.09}$ when adding  $H(z)$ data at 95\% confidence.

If we make the identification of $\tau\sim 2\epsilon z$, this represents an improvement of almost a factor 7 on the constraints  obtained in \cite{More} of  $\Delta\tau\equiv\tau(0.35)-\tau(0.2)<0.13$ at 95\% confidence: we obtain $\Delta\tau < 0.02$. This can be understood because they used only data at two redshifts to constrain $\Delta \tau$ in  a parameterization-independent way, while we use a range of redshifts to obtain a parameter-dependent constraint on $\tau(z)$. In both analyses one parameter is effectively measured, in the case of \cite{More} the parameter is $\Delta\tau$, while in our case it is $\epsilon$.

 Since our constraints are parametric, we should examine the extend to which
our results depend on the chosen parameterization.  Opacity should increase monotonically
with redshift, while the explanation of the observed behaviour of  supernovae brightness as function of redshift in terms of imperfect transparency would require a change of monotonicity at $z>1$ (supernovae get dimmer at $z<0.3$ and brighter at higher redshifts).  Thus, for generic parameterizations respecting the above monotonicity one expects to obtain strong constraints on cosmic transparency,
even though the quantitative details will depend on the particular parameterization.
Further, since the $H(z)$ data are not sensitive to opacity, the combined data-set constraints
will be even less sensitive.

For example for $z<0.5$ we find that $\epsilon=0.9$ mimics to better than 4\%  the replenishing dust model shown by \cite{Riess04} to explain Supernovae results without resorting to cosmic acceleration. Still, Fig. 1 shows that the combined supernovae+$H(z)$ constraints do not degrade significantly.
To demonstrate this further we consider an alternative parametrization,
which is linear in cosmic time rather than redshift, and which attempts to  model supernova evolution. In this model the apparent Supernovae dimming is given
by \cite{ferramacho}:
\be
\Delta m = K \frac{t_0-t}{t_0-t_1} \,.
\ee
Here $t_0$, $t_1$ is cosmic time at $z=0$, $1$ respectively, while $K$ is the
``opacity" parameter to be constrained. In this case we find the constraints to be  similar to  the dark 
background contours of  Fig. 1: empirically  we find that there is a linear  correpondance between $\epsilon$ and  $K$ given by  $\epsilon= 0.65 K+0.03$ that produces the same contours in Fig 1. One thus obtains from the joint analysis with $H(z)$ data $K< 0.3$.

Further, one may worry that the Taylor expansion we used to connect our constraints on
$\epsilon$ to constraints on $\tau$ may break down.  We have verified that the expansion
is good to $20\%$ for the entire range of $\epsilon$ and the redshift range considered.
The maximum deviation occurs for $\epsilon<-0.4$ at large redshifts $z\simeq 1.5$, while at $\epsilon\simeq -0.08$ (constraint obtained from SN data only), and $\epsilon\simeq -0.01$ (constraint from the combined SN and H(z) data),
the only parameter regions  where we have interpreted $\epsilon$ in terms of opacity $\tau$,
the expansion is accurate to better than $10\%$ and $1\%$ respectively, even at the maximum redshift considered.

\section{Consistency with the measured  BAO scale}
\begin{figure}[h]
  \begin{center}
    \includegraphics[width=3.5in,keepaspectratio]{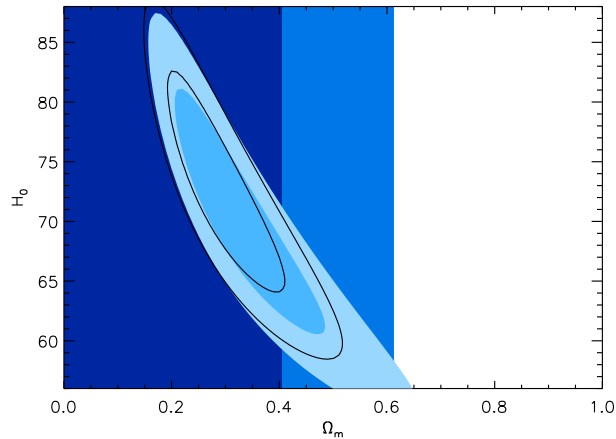}
  \end{center}
  \caption{\label{H0_omm_constr} Likelihood contours for flat $\Lambda CDM$ models
  in the $\Omega_m$-$H_0$ plane. Dark blue regions show the Supernovae constraints (data from all redshifts),  while the dark blue ones are for $H(z)$ constraints. Black contours show the joint
  constraints. By construction supernovae data  do not constrain $H_0$.}
 \end{figure}

Having  allowed the Supernovae data the extra degree of freedom of violating photon conservation, we can  check the cosmological constraints one can obtain  in combination with $H(z)$ data.
This is shown in figure \ref{H0_omm_constr}, where the dark blue regions are the Supernovae constraints, the light blue ones  are for $H(z)$ and the black contours are the joint constraints. Note that  as seen before the Supernovae data only give an upper limit to $\Omega_m$ and, by construction, supernovae data do not constrain $H_0$. These constraints are in good agreement with the WMAP5-determined parameters \cite{wmap5}.

A hint of  a discrepancy at the $2.4$-$\sigma$ level between Supernovae and BAO data at redshifts $z=0.2$ and $z=0.35$ was indicated in
\cite{PercivalBAO}.
 On the other hand,  \cite{EisensteinBAO} find agreement for BAO data at $z=0.35$.  To investigate whether this  discrepancy can be explained by a non-zero $\epsilon$,
we set out to compare these measurements of the BAO scale with our distance measurements of \S 1.

 BAO surveys  in principle could measure independently $H(z)$ and $d_A(z)$, or, more precisely, $d_A(z)/r_s$ and $H(z) r_s$ from transversal and radial clustering respectively.  Here $r_s$ denotes the sound horizon at radiation drag. Current surveys, however,  can only measure an angle-averaged distance measure $D_V(z)$ in combination with $r_s$.
\be\label{Dv}
D_V(z)=\left(\frac{cz(1+z^2)d_{\rm A}(z)^2}{H(z)}\right)^{1/3} \,.
\ee
The sound horizon $r_s$ is determined exquisitely by CMB data so it is possible to translate constraints on $r_s/D_V$ into constraints on $D_V$ and vice versa without degrading the error-bars.
In particular, \cite{EisensteinBAO} reports constraints on $D_V(0.35)$ obtained from SDSS Luminous red galaxies, while \cite{PercivalBAO} reports joint constraints on $r_s/D_V(0.2)$ and $r_s/D_V(0.35)$ from a combination of the SDSS and 2dFGRS surveys. On the CMB side,  WMAP 5-year data analysis \cite{wmap5} reports constraints on $r_s$, $r_s/D_V(0.2)$, $r_s/D_V(0.35)$.  Finally, our joint analysis of Supernovae and $H(z)$ (Fig.\ref{H0_omm_constr}) also yield joint constraints on $D_V(0.2)$ and $D_V(0.35)$.

\begin{figure}[h]
  \begin{center}
    \includegraphics[width=0.8\columnwidth,keepaspectratio]{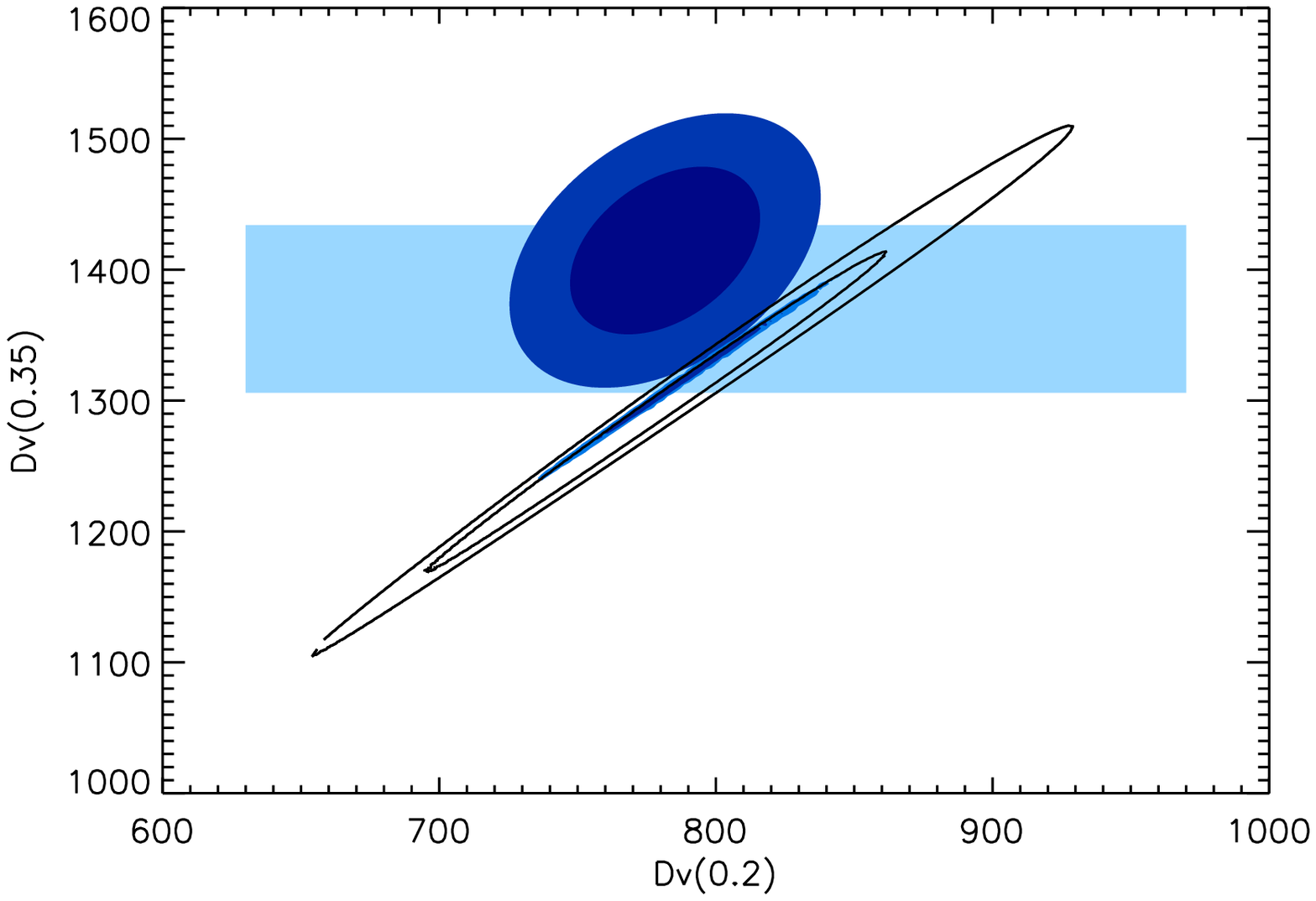}
    \includegraphics[width=0.8\columnwidth,keepaspectratio]{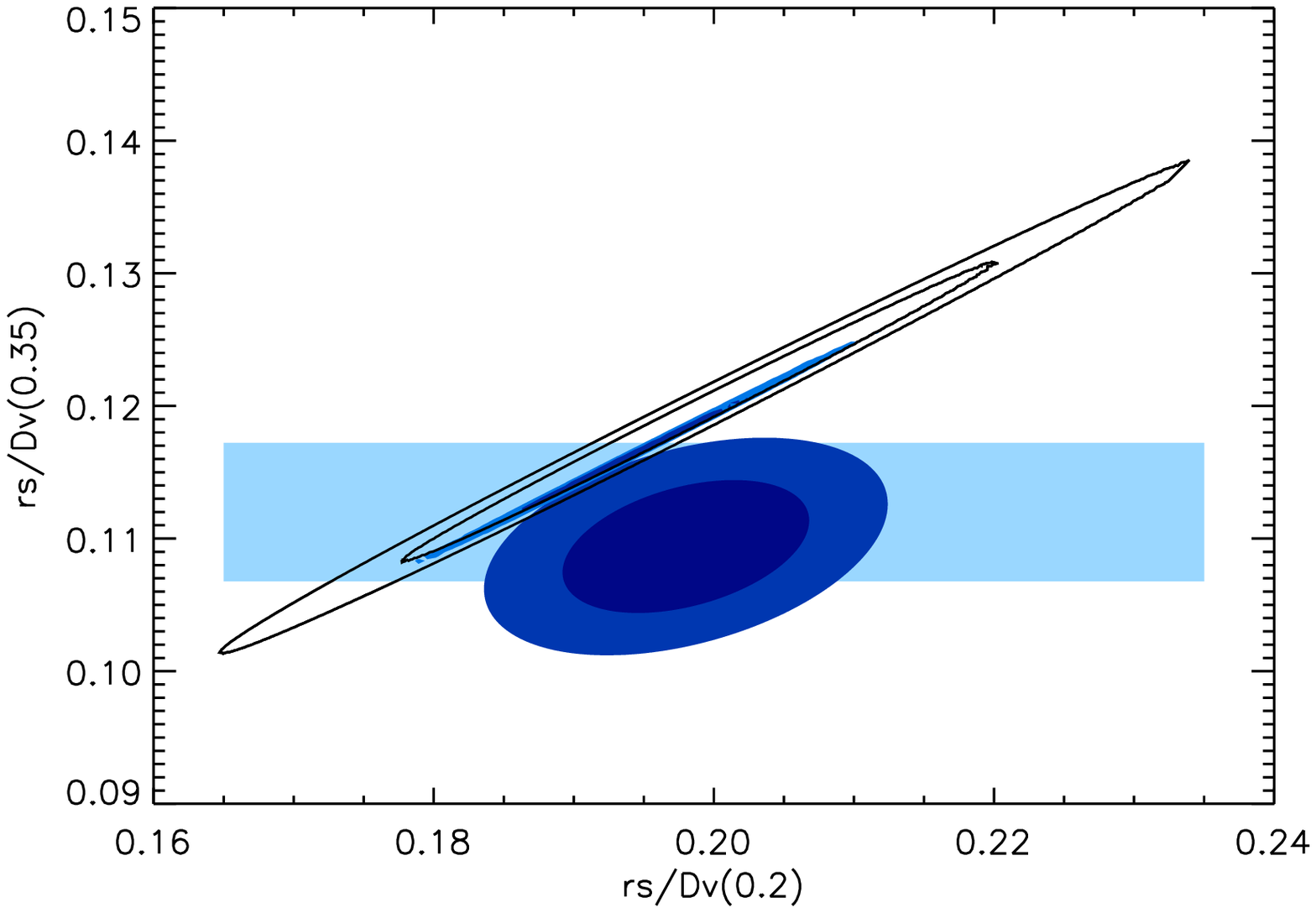}
  \end{center}
  \caption{\label{Dv02Dv035_alldata}  Constraints from CMB, BAO, and Supernovae+$H(z)$ in the $D_V(0.2)$, $D_V(0.35)$ plane (top panel) and $r_s/D_V(0.2)$, $r_s/D_V(0.35)$ (bottom panel). The underlying model is assumed to be a flat $\Lambda$CDM. The transparent contours are the constraints  obtained from Supernovae and $H(z)$ data  allowing for non-zero $\epsilon$,  the thin filled contours are the WMAP5-years constraints ($1$ and $2$-$\sigma$ joint), the light blue region corresponds to  the   $1$-$\sigma$ constraint from \cite{EisensteinBAO} (constrains only $D_V$ at $z=0.35$), and the dark blue confidence regions correspond to  the $1$ and $2$-$\sigma$ joint constraints from \cite{PercivalBAO}. }
 \end{figure}

In figure \ref{Dv02Dv035_alldata} we show constraints from these data sets separately in the $D_V(0.2)$, $D_V(0.35)$ plane (top panel) and $r_s/D_V(0.2)$, $r_s/D_V(0.35)$ (bottom panel). As before, the underlying model is assumed to be a flat $\Lambda$CDM. In this figure the transparent contours are the constraints obtained from Supernovae and $H(z)$ data  allowing for non-zero $\epsilon$ (corresponding to the solid black lines of Fig. \ref{H0_omm_constr}). The thin filled contours are the WMAP5-years constraints. The light blue region corresponds to the $1$-$\sigma$ constraint from \cite{EisensteinBAO} (note that \cite{EisensteinBAO} constrains only $D_V$ at $z=0.35$), and the dark blue confidence regions correspond to the $1$ and $2$-$\sigma$ joint constraints from \cite{PercivalBAO}.

Figure \ref{Dv02Dv035_alldata} shows that imperfect transparency cannot explain the tension between BAO and Supernovae distances. When allowing for imperfect transparency, our joint SN-H(z) constraints give rise to much broader contours than the corresponding WMAP constraints, but  the reported tension remains  even with this different data set.  Supernovae distances are consistent with both CMB data and high-redshift BAO constraints, and only the constraints of  \cite{PercivalBAO}  show some tension.
In interpreting this figure one should keep in mind that the data of \cite{PercivalBAO} and \cite{EisensteinBAO} are not fully independent: the data at $z=0.35$ are largely common to the two analyses while the error-bars include statistical errors. In agreement with \cite{PercivalBAO} we see that the discrepancy from the $\Lambda$CDM model comes mainly  from the low redshift BAO distance measure.  The value of $D_V(0.2)$ appears to be  too low by 5 to 10\%.
Even considering only  the Supernovae and BAO data of \cite{PercivalBAO},  an imperfect transparency ($\epsilon>0$) would make the discrepancy worse,  as Supernovae would favor less acceleration while the low redshift BAO data  require stronger acceleration. An $\epsilon<0$ would achieve this, however,  it cannot be interpreted in terms of reduced transparency, and the addition of $H(z)$ data  puts a strong lower limit on $\epsilon$, excluding this possibility.
It is well known that  non linearities, (increasingly more important at low redshift), may shift the  observed position of the BAO as traced by galaxies  (e.g., \cite{SeoSiegelEisensteinWhite} and references therein)     and it does so in the direction of reducing the inferred $D_V$. However, this effect is bound to be at, or below, the 1\% level for the robust technique to recover the signal used by \cite{PercivalBAO}; it cannot therefore be solely responsible for the shift in the measured $D_V(0.2)$.
Figure \ref{Dv02Dv035_alldata}  and these considerations seem to indicate that the source of the discrepancy may most likely be found among systematic effects of the modelling of the low redshift data or a simple $\sim 2$-$\sigma$ statistical fluke, rather than in exotic physics.

\section{Model-independent constraints on acceleration}

\begin{figure}[h]
  \begin{center}
   \includegraphics[width=3.5in,keepaspectratio]{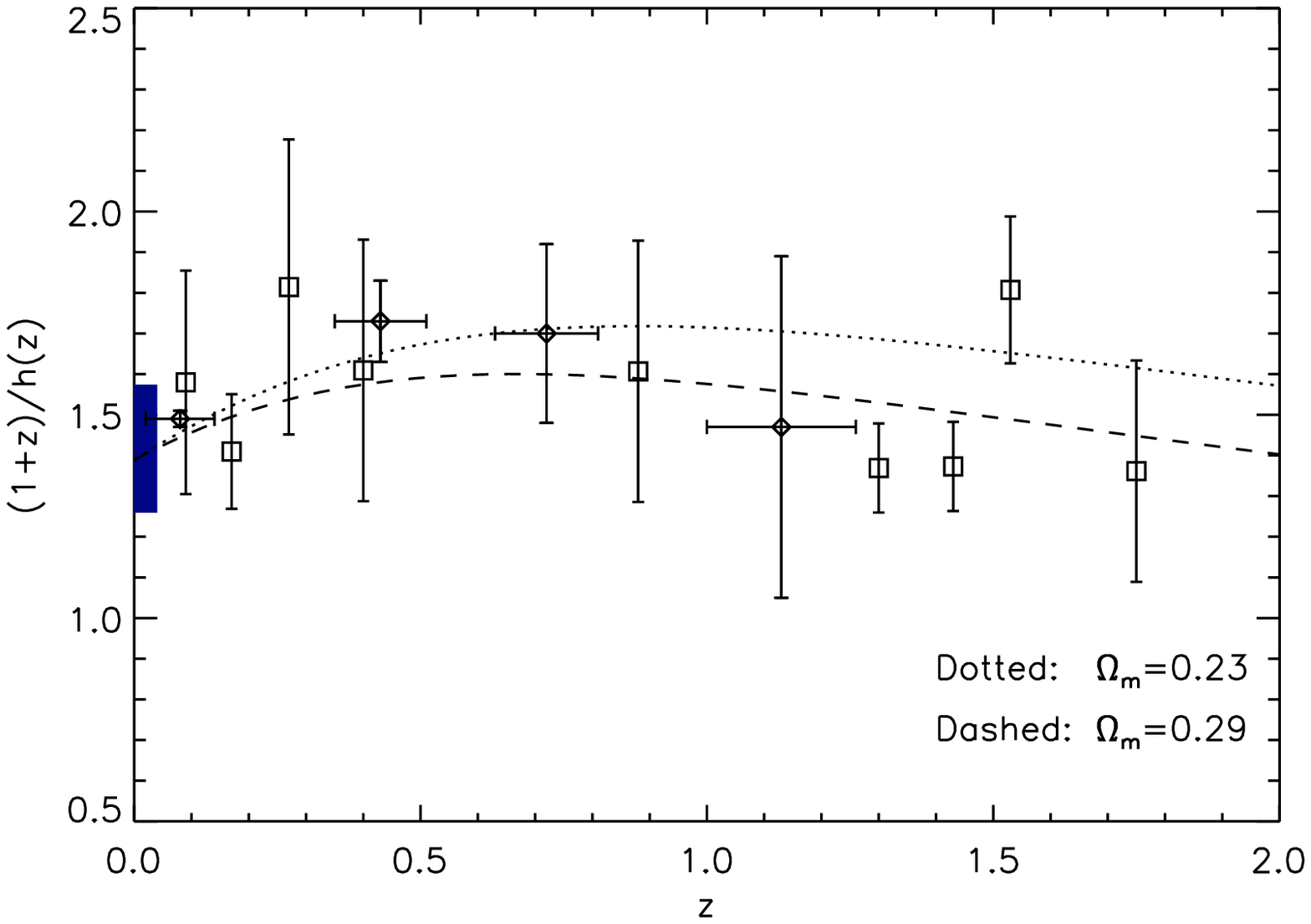}
    \includegraphics[width=3.5in,keepaspectratio]{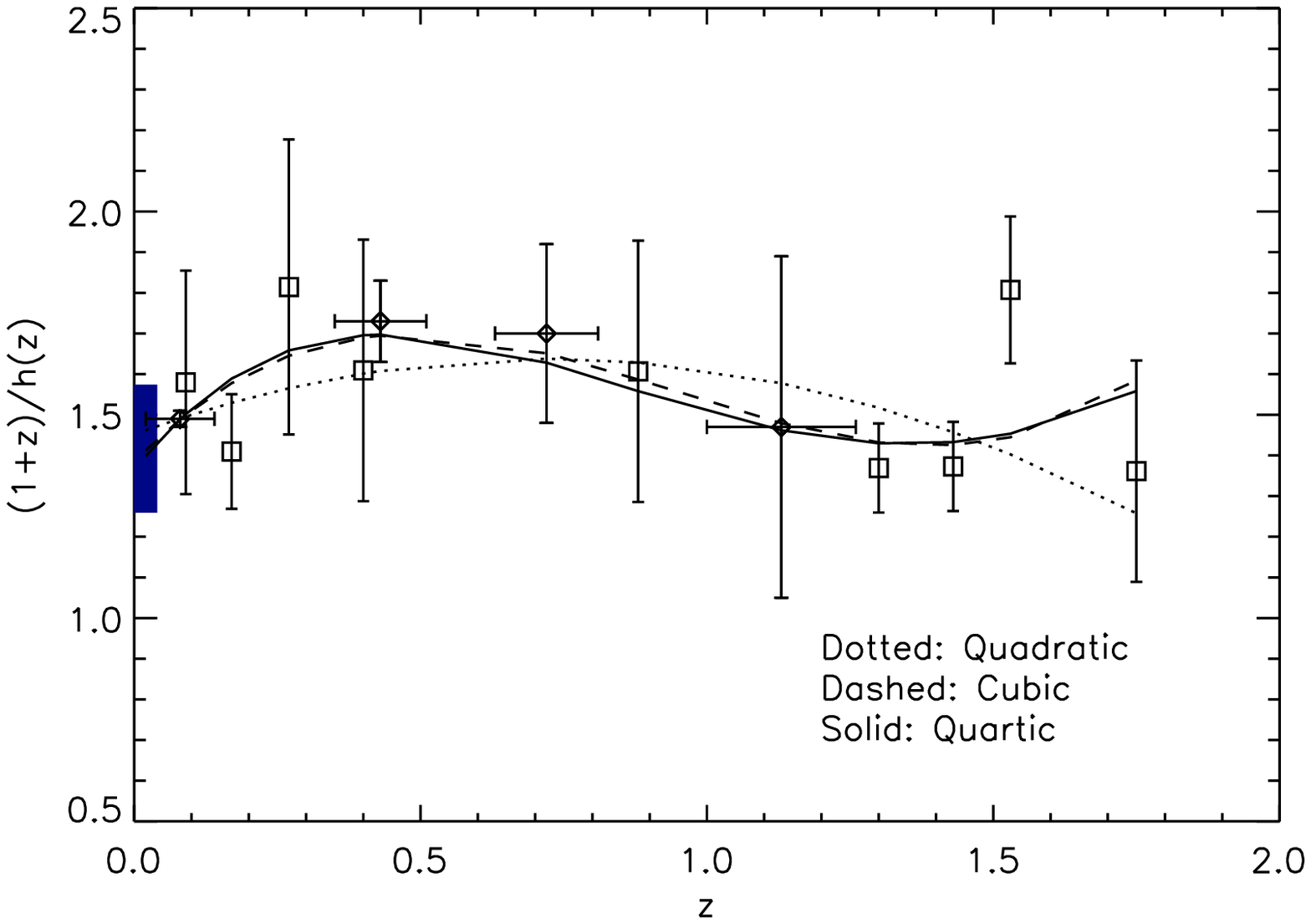}
  \end{center}
  \caption{\label{polyfits} {\bf Top:} Our data set combination:  HST Key project determination of $H_0$ (filled square), points from Supernovae~\cite{MortsellClarkson} (filled dots), points from $H(z)$ determinations of \cite{SVJ05}. Note $h(z)\equiv H(z)/100$. For comparison two $\Lambda$CDM model lines are plotted. {\bf Bottom:} Quadratic, cubic and quartic polynomial fits.}
\end{figure}

Having explored the complementarity and consistency of different distance measures and having found that Supernovae and $H(z)$ data are consistent, we now attempt to draw some model-independent conclusions about  the recent  accelerated expansion by combining them.
There is evidence for recent acceleration at  high-statistical significance, but most analyses in the literature are done in a model-dependent, parametric way, where the  acceleration is due to a dark energy component with properties specified by few parameters (e.g., $\Omega_V$, $w$ or $\Omega_V$, $w_o$, $w_a$). A fully non-parametric analysis is probably impossible, but we can attempt a model-independent  approach.

To do so we follow \cite{MortsellClarkson, WangTegmark,Riessetal07} and  introduce the function $f(z)= (1+z)/H(z)=\dot{a}^{-1}$: $f(z)$ has positive slope  if the universe is accelerating and negative if it is decelerating. In fact the deceleration parameter $q=-\ddot a a/\dot a^2$ is related to $f(z)$ by $q=-H(z) f^{\prime}(z)$.

While the $H(z)$ data of \cite{SVJ05} and \cite{HSTKey} for $z=0$ yield $f(z)$ directly, Supernovae-determined $d_L$ does not, and $f(z)$ can be inferred only taking a derivative \cite{Riessetal07,WangTegmark,Daly}.  \cite{MortsellClarkson} presented  such a model-independent  analysis  and in particular provided  estimates for $f(z)$ from the same Supernovae data set used here, which we now combine with HST-key \cite{HSTKey} and determinations of $H(z)$ \cite{SVJ05}. The details of obtaining the expansion history $H(z)$ by differentiating Supernovae data can be
found in Refs.~\cite{WangTegmark,MortsellClarkson}, the method assumes a spatially flat universe so we will only consider the zero-curvature case.

 The resulting data points are shown in the top panel of Fig. \ref{polyfits}, where two $\Lambda$CDM models are also shown for comparison.  A visual inspection of Fig.~\ref{polyfits} may indicate that the data points follow the expected trend (acceleration at low $z$ and deceleration at large $z$), but that given the large errors in $f(z)$, the data may also be consistent with a constant  function of $z$ (no acceleration/deceleration).

To investigate the statistical significance of  a possible deviation from a constant $f(z)$, we consider
three different analyses, namely: {\it i)} weighted fits with linear, quadratic, cubic and quartic polynomials,  {\it ii)} weighted linear fits up to different upper redshifts, and {\it iii)} piecewise linear fits. These should be seen as model-independent parameterizations of  the behavior of $f(z)$. We use more than one parameterization to assess the robustness of our findings:

 Weighted fits with quadratic, cubic and quartic polynomials yield curves that are increasing --corresponding to acceleration-- up to  redshift $z_{acc}=0.75, 0.45, 0.45$ respectively  and decreasing --corresponding to deceleration-- at $z>z_{acc}$, although at $z> 1.5$ the reconstruction become noisy. This is shown in the bottom
 panel of Fig. \ref{polyfits}.
 The best fit  chisquares are  10.59, 7.22, 7.17
 quadratic (3 parameters), cubic (4 parameters) and quartic (5 parameters) polynomials respectively.  The linear fit  has a best fit chisquare of  13.93. As there are only 14 data points, a simple chisquare  analysis indicates that  a straight line is a good fit and that there is no evidence for acceleration or deceleration. Nevertheless, one should consider the possibility that the errors may be over-estimated.
We thus attempt  to treat the linear and polynomial fits as nested models and apply information criteria as model selection tool.


 The Akaike information criteria
 (AIC) for the linear, quadratic, cubic and quartic fits are 17.9, 16.59, 15.22 and 17.17 respectively, while
 the corresponding Bayesian information criteria (BIC) are 19.2, 18.5, 17.77 and 20.36.  Both AIC and BIC favour  the cubic fit.  In particular, the BIC provides ``positive evidence" (in the Jeffreys scale~\cite{Jeffreys})  in favour of the cubic polynomial with respect to the linear model, with a Jeffreys factor of $\Delta$(BIC)$\sim 2$.  We interpret this as  slight positive evidence for model-independent detection of recent cosmic  acceleration at redshifts smaller than $0.5$ or so, and past deceleration at large redshifts ($z>1$).
 It is worth keeping in mind  that although both relative measures favour the cubic fit, the reduced chisquared for the linear model is close to unity  suggesting that the linear model is also a good fit to the data.  We interpret this as indication that the  error bars for the used measurements may have been overestimated.

\begin{figure}[h]
  \begin{center}
   \includegraphics[width=3.5in,keepaspectratio]{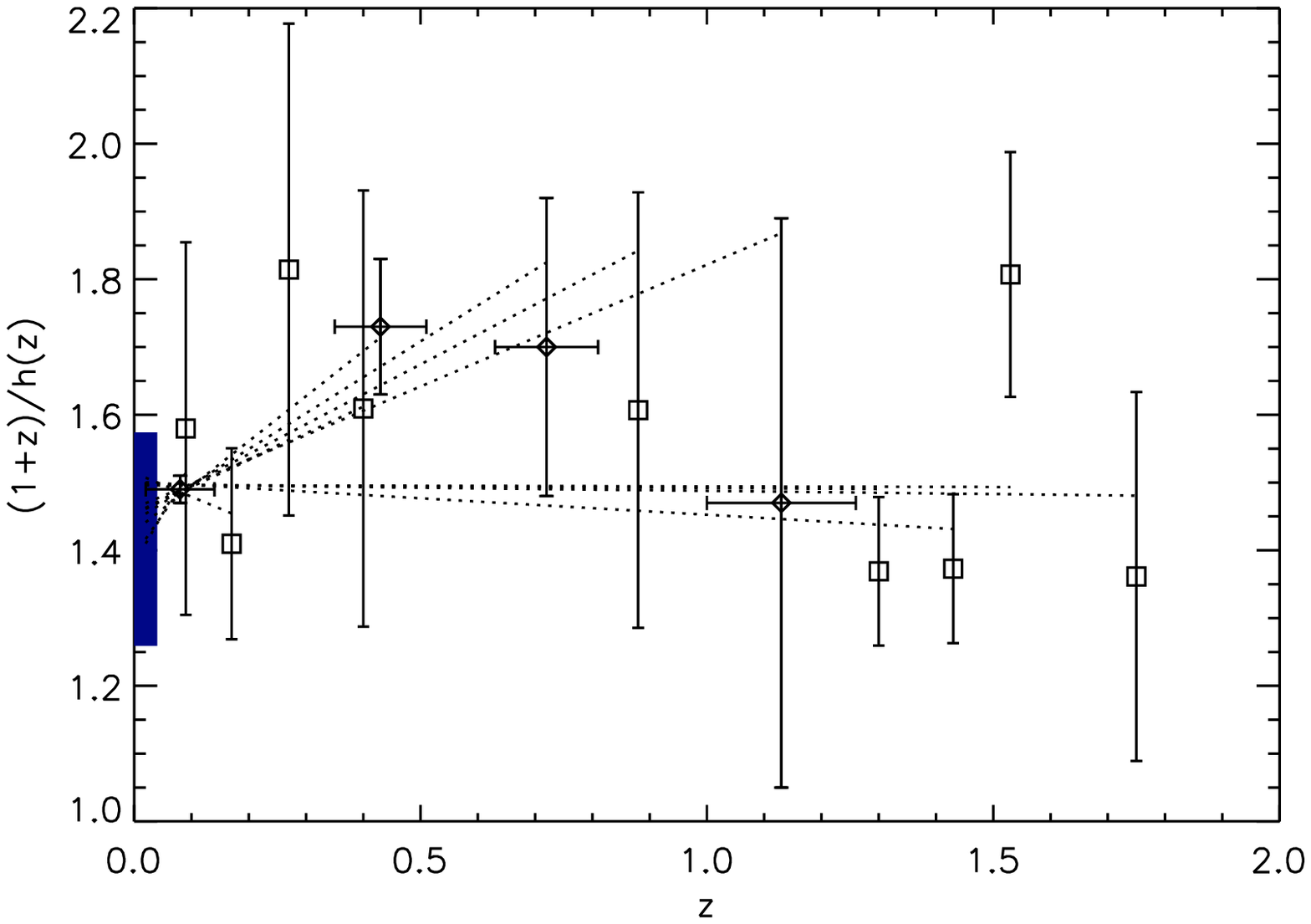}
    \includegraphics[width=3.5in,keepaspectratio]{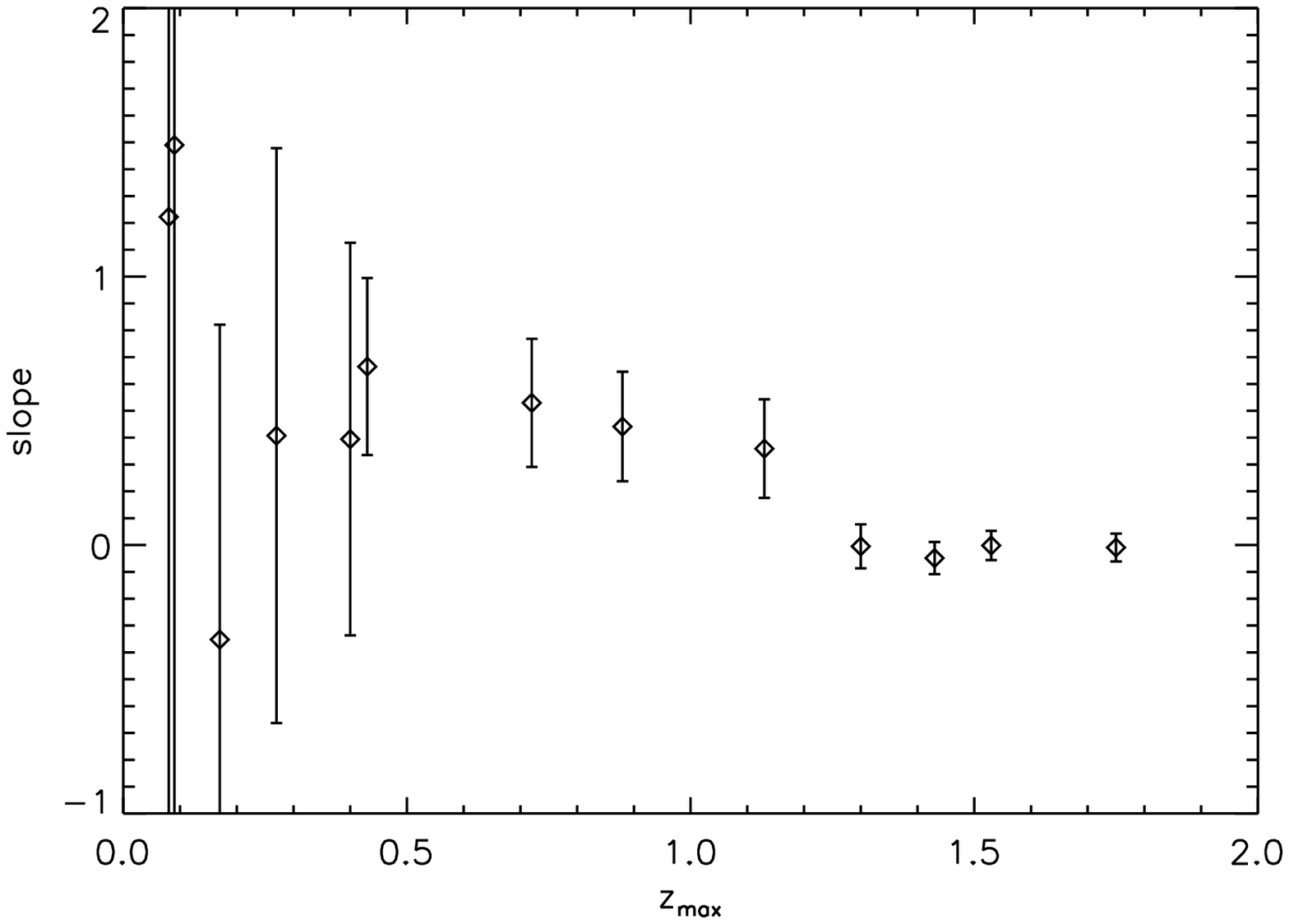}
  \end{center}
  \caption{\label{linfits} {\bf Top:} Linear fits up to redshift $z_{max}$.  Lines broadly fall into
          two categories, lines with a positive slope of $0.5\pm 0.15$ ($z_{max}<1$), and lines with flat
           slopes for $z_{max}>1.1$. {\bf Bottom:} Slope as a function of $z_{max}$. Starting
           from left and moving towards higher $z_{max}$, more points are included in the fitting.}
\end{figure}

We then try to understand better the result of the linear fit model by  considering cases {\it ii)} and {\it iii)}: weighted linear fits up to different upper redshifts, and  piecewise linear fits.
The weighted linear fits as a function of the maximum redshift considered, $z_{max}$, are shown in Fig.~\ref{linfits}:  lines with $z_{max}$
between 0.3 and 1, have a positive slope, and  lines with higher $z_{max}$
have  flat or slightly negative slopes.  Fig.~\ref{linfits} (bottom panel) shows
the slopes of the lines (with the corresponding $1$-$\sigma$ errors) as a function of $z_{max}$.
Thus, by fitting a single line from $z=0$ to $z_{max}$ we see evidence for acceleration at low $z$, but no clear indication of acceleration or deceleration when all redshifts  up to $z_{max}>1$  are considered.  This analysis
cannot provide an estimate for the redshift of acceleration onset.

\begin{figure}[h]
  \begin{center}
   \includegraphics[width=3.5in,keepaspectratio]{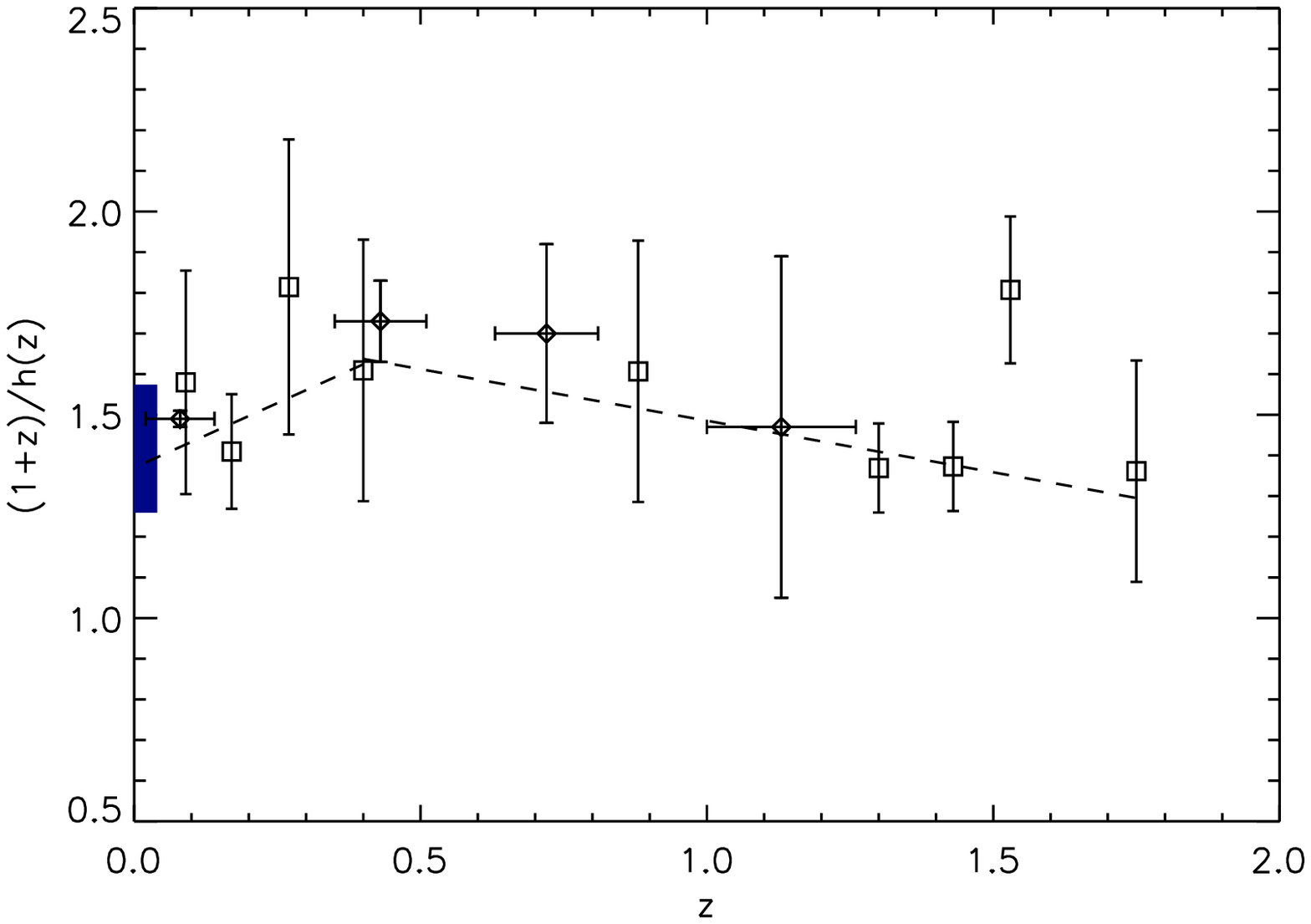}
    \includegraphics[width=3.5in,keepaspectratio]{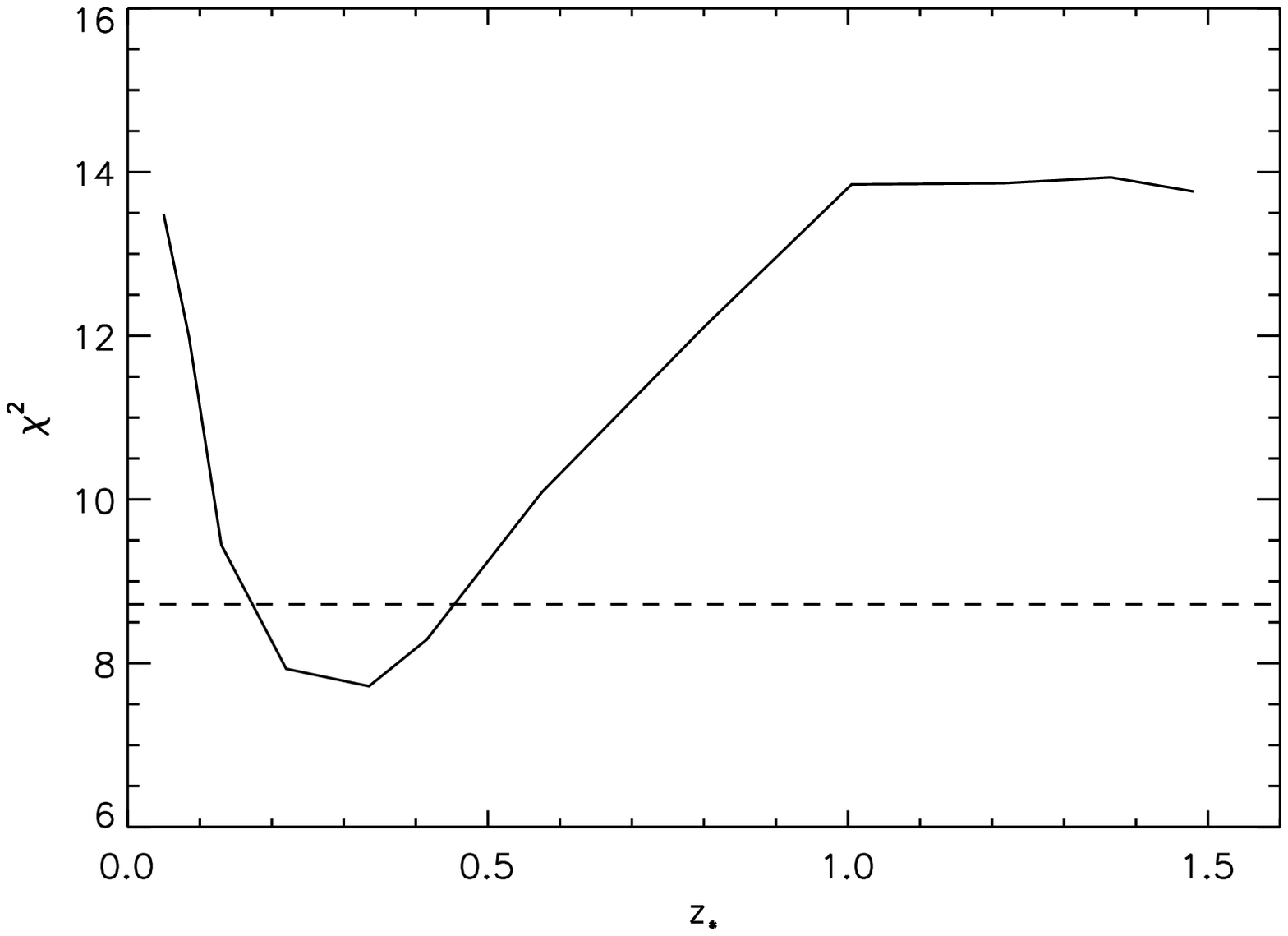}
  \end{center}
  \caption{\label{bilinfits} {\bf Top:} The best fit continuous piecewise linear fit.   {\bf Bottom:} Chisquare as a function of the transition redshift between deceleration and acceleration $z_{acc}$. Dashed line indicates a $\Delta \chi^2=1$.}
\end{figure}

A bilinear fit on the other hand, can give a model-independent estimate of the transition redshift $z_{acc}$.
 We perform a continuous piecewise linear fit  dividing the data in two redshift intervals: from redshift 0 to $z_*$ and
from $z_*$ to $z=1.8$.  In this way $z_*$ yields an estimate of the acceleration redshift $z_{acc}$.  As before, the reduced chisquares are  always  $< 1$: any $z_{acc}$ gives formally an acceptable fit. However, the chisquare changes significantly as a function of the parameters.  The top panel of  figure \ref{bilinfits} shows  the best fit (acceleration slope $df/dz=0.63\pm 0.18$ and deceleration slope $df/dz=-0.35 \pm 0.08$ for the best fit $z_{acc}=0.35$). The bottom panel shows  the
chisquare as a function of  $z_{acc}$. Note that  while goodness of fit tests considers the absolute value of  $\chi^2$ and all values of  $z_{acc}$ pass the goodness of fit test, parameter estimation procedure only considers $\chi^2$ {\it differences} from the minimum. The bottom panel of Fig. \ref{bilinfits} clearly shows that  when looking at $\Delta \chi^2$, the data  constrain $z_{acc}$; in particular, $z_{acc}=0.35^{+0.20}_{-0.13}$  at 1-$\sigma$ level. This may indicate that the error-bars may be over-estimated or highly correlated.

 The above analysis has been carried out without reference to any specific model of the energy
content of the universe, nor to any specific theory of gravity.  We have assumed, however, homogeneity
and isotropy at large scales, so that the FLRW geometry can be considered a good approximation.  In
addition, the method used to obtain four out of the fourteen data-points used in this section (those coming from differentiating Supernovae data) assumes spatial flatness.

\section{Conclusions}

Combining different distance measurements is a powerful tool to constrain cosmological parameters. Comparing them instead helps  constrain possible deviations from the assumptions underlying the standard  cosmological model or exotic physics. We have considered Supernovae data~\cite{Union}, direct meauserments of cosmic expansion~\cite{SVJ05,HSTKey} and  determinations of the
BAO scale ~\cite{PercivalBAO,EisensteinBAO}.

We have explored the possibility of uniform deviations from ``cosmic transparency" through its
effects on distance duality, by parameterizing possible deviations from the Etherington
relation (\ref{Etherington}). Such deviations from the  Etherington  relation might arise from any  uniform and ``gray" source of attenuation, from astrophysical processes such as  gray \cite{Aguirre} or replenishing dust \cite{Riess04} (used to try to explain Supernovae results without resorting to cosmic acceleration, or relaxing the constraints on cosmic acceleration),  to more exotic  physics such as photon-axion mixing \cite{Csaki1,Csaki2,BassettKunz1,Burrage}.

Combining direct meauserments of cosmic expansion~\cite{SVJ05}
and SN data~\cite{Union}, and assuming an underlying flat $\Lambda$CDM
model, we have placed strong constraints on such deviations, in
particular improving the existing constraint~\cite{More} by nearly an order of magnitude.

While allowing for such deviations from transparency, we have revisited the reported
tension~\cite{PercivalBAO} between distance measures obtained from SN and
BAO within the $\Lambda$CDM model.
We  have found no discrepancy among   Supernova data --even allowing for deviations from transparency-- , $H(z)$ determinations  and  high-redshift BAO, indicating that
the source of the discrepancy may likely be found among systematic effects in the
modelling of low redshift  BAO data (or be attributed to a $2$-$\sigma$ statistical fluctuation),
rather than in exotic physics.

Finally, since Supernovae data and $H(z)$ data appear to be  highly consistent with each other, we have combined them and attempted to draw some model-independent conclusions about the recent accelerated expansion. We determine the transition redshift between deceleration and recent acceleration to be $z_{acc}=0.35^{+0.20}_{-0.13}$.

 To conclude, we remark that the comparison among  distance measurements along the lines of the  analysis presented here can also be used to place bounds on possible departures
from homogeneity and isotropy.  There has been significant activity recently in trying to use comparisons among  distance measurements as a powerful tool for probing fundamental assumptions behind cosmological models.  For example,
Clarkson et al~\cite{Clarkson} suggest to test the Copernican Principle by examining the consistency condition:
$ {\cal C}(z)\equiv 1+H^2(DD^{\prime\prime}-{D^{\prime}}^2)+HH^{\prime}DD^{\prime}=0 \, ,$
where $D=(1+z)^{-1}d_L$ and a prime stands for differentiation with respect to redshift
$z$.  This condition arises by solving for the curvature energy density parameter $\Omega_k$
from the definition of the luminosity distance in a FLRW universe, and then demanding that its
derivative vanishes.  In general, deviations form the Copernican Principle may appear as inconsistency between distance measures. The condition  above  relies on the same consistency relation we used here, that is
on the consistency relation that exists within the homogeneous and isotropic FLRW models between luminosity or area distance and the Hubble rate, both as a function of redshift.  In particular  the Clarkson et al condition is not satisfied for radially inhomogeneous models in general.

While our data are too noisy to perform the test in the form stated above (note
that ${\cal C}(z)$ involves first and second derivatives of the data),  here, in the language of \cite{Clarkson}, we have  effectively looked at radial deviations from the Copernican Principle by testing consistency between different distance measurements ($d_L(z)$ and $H(z)$).
We  have found no  indication of inconsistency among the different datasets/distance measures.
In general, radial inhomogeneities would lead to anisotropic clustering,
so, for example, the BAO feature in the two-point correlation function would not be isotropic~\cite{Clarkson}, yielding inconsistency between the  inferred $d_A$ and $H(z)$. A similar effect is expected in a large class of Bianchi models.   We expect that future data will impose interesting constraints on these models.

\section*{Acknowledgements}
AA would like to thank B. Reid, C. Carbone, J. Miralda-Escud\'e and J. Garriga for useful discussions.  AA is
supported by the Institute of Space Sciences (IEEC-CSIC).  LV is supported by FP7-PEOPLE-2007-4-3 IRG n 202182,
RJ is supported by FP7-PEOPLE-2007-4-3 IRG. LV and RJ are supported by MICINN grant AYA2008-03531.
We thank the University of Barcelona and in particular the Institute of Cosmos Sciences for hospitality.
We also thank the Galileo Galilei Institute for Theoretical Physics for hospitality and INFN for partial
support during the completion of this work.  We acknowledge the use of the Supernova Cosmology Project (SCP)
``Union'' Cosmology Tables and of the Legacy Archive for Microwave Background Data Analysis (LAMBDA).  Support
for LAMBDA is provided by the NASA Office of Space Science.

\end{document}